\documentstyle[amstex]{mn}
\newcommand{\apj}{Astrophys. J.}

\input epsf
\def\plotone#1{\centering \leavevmode
\epsfxsize=\columnwidth \epsfbox{#1}}

\def\plottwo#1#2{\centering \leavevmode
\epsfxsize=.99\columnwidth \epsfbox{#1} \hfil
\epsfxsize=.99\columnwidth \epsfbox{#2}}

\def\plotthree#1#2#3{\centering \leavevmode
\epsfxsize=.99\columnwidth \epsfbox{#1} \hfil
\epsfxsize=.99\columnwidth \epsfbox{#2} \hfil
\epsfxsize=.99\columnwidth \epsfbox{#3}}

\newif\ifAMStwofonts

\ifoldfss
  \ifCUPmtlplainloaded \else
    \NewTextAlphabet{textbfit} {cmbxti10} {}
    \NewTextAlphabet{textbfss} {cmssbx10} {}
    \NewMathAlphabet{mathbfit} {cmbxti10} {} 
    \NewMathAlphabet{mathbfss} {cmssbx10} {} 
  \fi
  \ifAMStwofonts
    \ifCUPmtlplainloaded \else
      \NewSymbolFont{upmath} {eurm10}
      \NewSymbolFont{AMSa} {msam10}
      \NewMathSymbol{\upi}     {0}{upmath}{19}
      \NewMathSymbol{\umu}     {0}{upmath}{16}
      \NewMathSymbol{\upartial}{0}{upmath}{40}
      \NewMathSymbol{\leqslant}{3}{AMSa}{36}
      \NewMathSymbol{\geqslant}{3}{AMSa}{3E}

    \fi
  \fi
\fi 

\ifnfssone
  \newmathalphabet{\mathit}
  \addtoversion{normal}{\mathit}{cmr}{m}{it}
  \addtoversion{bold}{\mathit}{cmr}{bx}{it}
  \newmathalphabet{\mathbfit} 
  \addtoversion{normal}{\mathbfit}{cmr}{bx}{it}
  \addtoversion{bold}{\mathbfit}{cmr}{bx}{it}
  \newmathalphabet{\mathbfss} 
  \addtoversion{normal}{\mathbfss}{cmss}{bx}{n}
  \addtoversion{bold}{\mathbfss}{cmss}{bx}{n}
  \ifAMStwofonts
    \ifCUPmtlplainloaded \else
      \UseAMStwoboldmath
      \makeatletter
      \new@mathgroup\upmath@group
      \define@mathgroup\mv@normal\upmath@group{eur}{m}{n}
      \define@mathgroup\mv@bold\upmath@group{eur}{b}{n}
      \edef\UPM{\hexnumber\upmath@group}
      \new@mathgroup\amsa@group
      \define@mathgroup\mv@normal\amsa@group{msa}{m}{n}
      \define@mathgroup\mv@bold\amsa@group{msa}{m}{n}
      \edef\AMSa{\hexnumber\amsa@group}
      \makeatother
      \mathchardef\upi="0\UPM19
      \mathchardef\umu="0\UPM16
      \mathchardef\upartial="0\UPM40
      \mathchardef\leqslant="3\AMSa36
      \mathchardef\geqslant="3\AMSa3E
    \fi
  \fi
\fi 

\ifnfsstwo
  \DeclareMathAlphabet{\mathbfit}{OT1}{cmr}{bx}{it}
  \SetMathAlphabet\mathbfit{bold}{OT1}{cmr}{bx}{it}
  \DeclareMathAlphabet{\mathbfss}{OT1}{cmss}{bx}{n}
  \SetMathAlphabet\mathbfss{bold}{OT1}{cmss}{bx}{n}
  \ifAMStwofonts
    \ifCUPmtlplainloaded \else
      \DeclareSymbolFont{UPM}{U}{eur}{m}{n}
      \SetSymbolFont{UPM}{bold}{U}{eur}{b}{n}
      \DeclareSymbolFont{AMSa}{U}{msa}{m}{n}
      \DeclareMathSymbol{\upi}{0}{UPM}{"19}
      \DeclareMathSymbol{\umu}{0}{UPM}{"16}
      \DeclareMathSymbol{\upartial}{0}{UPM}{"40}
      \DeclareMathSymbol{\leqslant}{3}{AMSa}{"36}
      \DeclareMathSymbol{\geqslant}{3}{AMSa}{"3E}
    \fi
  \fi
\fi 

\ifCUPmtlplainloaded \else
  \ifAMStwofonts \else 
    \def\upi{\pi}
    \def\umu{\mu}
    \def\upartial{\partial}
  \fi
\fi

\title{Mixing through shear instabilities}
\author[]
       {M. Br\"uggen$^{1,2}$, W. Hillebrandt$^{1}$ \\
        $^1$ Max-Planck-Institut f\"ur Astrophysik,
Karl-Schwarzschild-Str.1, 85740 Garching, Germany\\ 
$^2$ Churchill College, Storey's Way, Cambridge, CB3 0DS, UK}

\pagerange{\pageref{firstpage}--\pageref{lastpage}}
\pubyear{1999}

\begin{document}

\maketitle

\begin{abstract}

In this paper we present the results of numerical simulations of the
Kelvin-Helmholtz instability in a stratified shear layer.  This shear
instability is believed to be responsible for extra mixing in
differentially rotating stellar interiors and is the prime candidate
to explain the abundance anomalies observed in many rotating stars.
All mixing prescriptions currently in use are based on
phenomenological and heuristic estimates whose validity is often
unclear.  Using three-dimensional numerical simulations, we study the
mixing efficiency as a function of the Richardson number and compare
our results with some semi-analytical formalisms of mixing.

\end{abstract}



\section{Introduction}

Mixing is a fundamental process which profoundly affects the evolution
of stars. Nonetheless, a theory of mixing that is applicable to
stellar conditions is still missing. Hitherto stellar evolution theory
has resorted to more or less heuristic prescriptions of mixing, the
validity of which is often unclear. One of the prominent mechanisms
responsible for mixing in stars is the shear or Kelvin-Helmholtz
instability. This instability is particularly important since most
stars are are known to be at least partly differentially
rotating. Therefore, shear mixing is likely to play a significant role
in stellar evolution.\\

Observational evidence seems to suggest that current prescriptions
underestimate the efficiency of the mixing processes at work,
especially in fast rotating stars. The observational evidence for
mixing is increasing rapidly. Herrero et al. (1992) find that all fast
rotating O-stars show significant surface He-enrichments. This seems
to suggest that mixing is strong enough to transport the nuclearly
processed material to the surface in a fraction of the life time of
the star. Other observations include the N/C and $^{13}$C and $^{12}$C
enrichments of the Red Giant Branch (e.g. Kraft 1997, Charbonnel
1995), the He and N excesses in OBA supergiants (Fransson et
al. 1989), the depletion of boron in most B-type stars (Venn, Lambert
\& Lemke 1996) and the ratio of the number of blue to red supergiants
in galaxies (Langer \& Maeder 1995). For a more detailed review of the
observational evidence see Maeder (1995), Kraft (1994) and references
therein. But the conclusion from these observations is that both, the
enrichment in CNO elements and the depletion of fragile elements such
as boron can be explained if some form of mixing is introduced (Langer
1997).\\

However, a convincing theory of mixing under most astrophysical
conditions is still missing. One of the pioneering works in this field
was performed by Endal \& Sophia (1978) who estimated the efficiency
of mixing due to several rotationally induced instabilities in massive
stars.

The Sun is also known to rotate differentially. Pinsonneault et
al. (1989) treated the efficiency of rotational mixing as a free
parameter and fitted it to a solar model. In the meantime,
helioseismological observations from the spacecraft SOHO and from the
GONG network have yielded precise measurements of the rotation rate in
the solar interior. In a number of publications the effects of mixing
on stellar evolution have been examined, see, e.g. Meynet \& Maeder
(1997) and Staritsin (1999). Recently, Denissenkov \& Tout (2000)
investigated the effects of `deep mixing' on the evolution of red
giants.  However, for want of a valid fundamental theory, the mixing
formalisms employed in the aforementioned studies have all been based
on more or less heuristic estimates. It is the aim of this work to
make steps towards a more fundamental theory of mixing. \\

In chemically homogeneous, stratified shear flows, the
Kelvin-Helmholtz instability occurs when the destabilising effect of
the relative motion in the different layers dominates over the
stabilising effect of buoyancy (see e.g. Chandrasekhar 1961).  The
competition between the two effects is described by the Richardson
number, Ri. For a non-dissipative, parallel, steady flow a simple
linear stability criterion can be derived as follows: Let the
horizontal velocity of the fluid be a function of $z$ only,
i.e. $U=U(z)$.  Now suppose that two volumes of fluid at heights $z$
and $z+\Delta z$ are interchanged where the densities at those two
heights are $\rho$ and $\rho +\Delta \rho$, and their horizontal
velocities $U$ and $U+\Delta U$, respectively.  The work per unit mass
that must be done against gravity is

\begin{equation}
\Delta W = -g\Delta\rho\Delta z .
\end{equation}
After the exchange both cells have an average velocity of
$\bar{U}=\frac{1}{2}(2U+\Delta U)$. Therefore, the kinetic energy
available to do this work is

\begin{equation}
{\textstyle\frac{1}{2}}\rho[U^2+(U+\Delta U)^2-{\textstyle\frac{1}{2}}(2U+\Delta U)^2] = {\textstyle\frac{1}{4}}\rho (\Delta U)^2 .
\end{equation}
Hence, a condition for stability is 

\begin{equation}
{\textstyle\frac{1}{4}}\rho (\Delta U)^2 < -g\Delta\rho\Delta z,
\end{equation}
or

\begin{equation}
\left (\frac{dU}{dz}\right )^2 < -4\frac{g}{\rho}\frac{d\rho}{dz}.
\end{equation}
A more general criterion can be derived (see e.g. Shu 1992):

\begin{equation}
{\rm Ri}:=\frac{N^2}{(dU/dz)^2} > \frac{1}{4},
\end{equation}
where $N$ is the Brunt-V\"ais\"al\"a frequency given by

\begin{equation}
N^2=\frac{g\delta}{H_P}\left [\left (\frac{\partial\ln T}{\partial\ln
P}\right )_{\rm ad}-\frac{\partial\ln T}{\partial\ln P} +\frac{\phi}{\delta}\frac{\partial\ln\mu}{\partial\ln P}\right ],
\end{equation}
with $\delta =-(\partial\ln\rho /\partial\ln T)_{P,\mu}$ and $\phi
=(\partial\ln\rho /\partial\ln \mu)_{P,T}$, where $T$ is temperature,
$P$ is pressure and $\mu$ chemical potential. $H_P$ is the pressure
scale height, i.e. $(\partial\ln P/\partial z)^{-1}$. Ri is the
so-called Richardson number. Stability is predicted if the {\it local}
Richardson number exceeds 1/4 everywhere.\\

More detailed analytical theories of shear mixing have been developed
(Maeder 1995, Canuto 1998) and some will be reviewed briefly in
Sec. 2. In Sec. 3, we present the results of three-dimensional
simulations of shear instabilities in stratified fluids.  These
numerical simulations were performed for a simple configuration and a
range of initial condition. Our aim is to elucidate the nonlinear
dynamics of the shear instability and to study the parametric
dependence of the mixing efficiency on the Richardson number. We
quantify the mixing of the fluid by introducing `tracer' particles
that are advected with the fluid. With conditions in stellar interiors
in mind, we only consider subsonic flows.  From the motion of these
particles a heuristic diffusion constant can be derived which is
then compared to the analytic prescriptions described in Sec. 2. \\

Apart from mixing by shear instabilities, there is another important
candidate believed to be reponsible for extra mixing in stars, namely
convective overshoot.  In models of stellar evolution the fluid is
assumed to be mixed whereever some convective instability criterion
(such as the Schwarzschild criterion) is fulfilled, and convectively
stable where this criterion is not fulfilled.  Mathematically the
boundary between these two regions is sharp.  In reality, however, the
convective fluid elements still carry some momentum when they reach
this boundary and subsequently penetrate into the convectively
``stable'' region.  This mechanism is called convective overshoot and
has been investigated numerically by a number of groups (Freytag,
Ludwig \& Steffen 1996, Nordlund \& Stein 1996, Singh, Roxburgh \&
Chan 1998) all of which find some degree of overshoot.  This
conclusion is supported by observations. The most compelling evidence
stems from isochrone fitting to stellar clusters and binary systems
which suggests that convective overshoot is significant (see Zahn 1991
for a review).
 
Unlike convective overshoot which provides mixing at the boundaries of
convection zones, we investigate mixing by rotational instabilities
that operate in convectively stable regions.

\section{Theories of shear mixing}

\subsection{Linear stability analysis of a shear layer}

The stability of a stratified shear layer has been studied extensively
using linear theory (see e.g. Chandrasekhar 1961, Howard \& Maslowe
1973, Maslowe \& Thompson 1971, Hazel 1972, Morris et al. 1990).
Recently, Balmforth \& Morrison (1998) reviewed the conditions for
instability in inviscid shear flows. In astrophysics, Kippenhahn \&
Thomas (1987) and MacDonald (1983) examined the redistribution of
matter and angular momentum in accreting white dwarfs by means of a
linear stability analysis. In this section we will present a simple
example of a linear stability analysis for a plane shear layer.  The
results of this analysis will then be compared against our numerical
simulations in Sec. 3.\\

\begin{figure}
\plotone{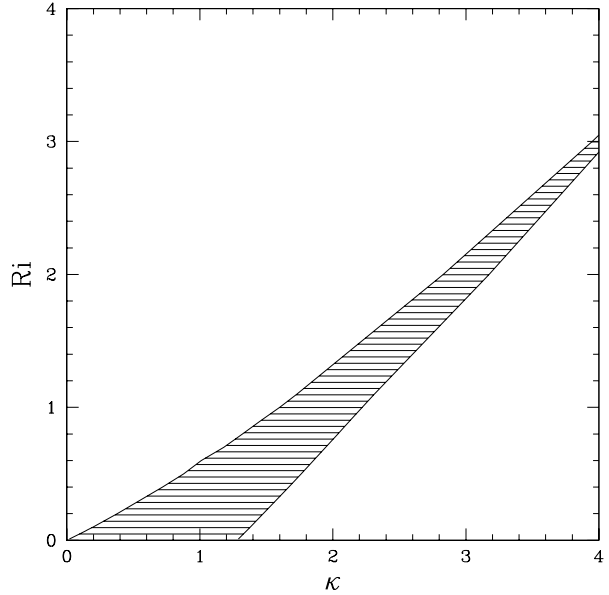}
\caption{Region of instability (shaded) in the $\kappa$-Ri plane, where
$\kappa$ is the dimensionless wave number and Ri the Richardson
number (see text).}
\end{figure}

The stability equation for wave-like perturbations in its inviscid and
incompressible form can be written as (e.g. Chandrasekhar 1961)

\begin{equation}
\frac{d^2\phi}{dz^2}-\left
(k^2+\frac{U''}{U-c}-\frac{g\beta}{(U-c)^2}\right
)\phi = 0 , \label{1.16}
\end{equation}
where $\beta = \rho '/\rho$ and $\phi (z)\exp [ik(x-ct)]$ is the
perturbation stream function.  Eqn. (\ref{1.16}) is sometimes referred
to as the Orr-Sommerfeld equation.  As an example for the linear
analysis of the Kelvin-Helmholtz instability we will cite a case which
was first considered by Taylor: It consits of two superposed fluids of different
densities separated by a transition layer of intermediate density in
which the shear velocity varies linearly from that of the lower layer
to that of the upper layer, i.e. \\

\noindent layer 1: $z>d$, $\rho = \rho_0(1-\epsilon)$,
$U=U_0$\\ layer 2: $d>z>-d$, $\rho = \rho_0$, $U=U_0z/d$\\ layer 3:
$z<-d$, $\rho = \rho_0(1+\epsilon)$, $U=-U_0$.\\

It is easily verified that the Richardson number for this problem is
given by

\begin{equation}
{\rm Ri}=\frac{g\epsilon d}{U_0^2} . 
\end{equation}
The solutions of eqn (\ref{1.16}) in each of the three regions are \\

\noindent layer 1: $\phi_1 = A_1 e^{-kz}$\\
layer 2: $\phi_2 = A_2 e^{-kz}+B_2 e^{kz}$\\
layer 3: $\phi_3 = A_3 e^{kz}$.\\

Continuity of $\phi$ at the interfaces, and requiring $\phi$ to
vanish at $\pm \infty$, gives the characteristic equation:

\begin{multline}
e^{-2\tilde{k}}=\left [1-\frac{\tilde{k}(\tilde{c}+1)^2}{{\rm
Ri}+(\tilde{c}+1)+\frac{1}{2}\epsilon\tilde{k}(\tilde{c}+1)^2}\right ]\\
\times \left [1-\frac{\tilde{k}(\tilde{c}-1)^2}{{\rm Ri}-(\tilde{c}-1)-\frac{1}{2}\epsilon\tilde{k}(\tilde{c}-1)^2}\right ]   ,  \label{char}
\end{multline}
where $\tilde{k}=2kd$ and $\tilde{c}=c/U_0$.  Eqn (\ref{char}) can now
be solved numerically for $c$ for a range of wave and Richardson
numbers. When $c$ becomes imaginary the perturbation becomes
unstable. The parameter region in which this occurs, i.e.  the region
of instability, is shown in Fig. 1.

The efficiency of mixing is sometimes parametrized by a diffusion
coefficient which can be loosely defined as 

\begin{equation}
D({\rm Ri}) \sim \frac{{\rm Im}\ [c_{\rm max}({\rm Ri})]}{k_{\rm max}({\rm Ri})} ,
\end{equation}
where ${\rm Im}\ [c_{\rm max}]$ is the maximal growth rate of the
perturbations and $k_{\rm max}$ the corresponding wave number - both
evaluated at a given Richardson number. The diffusion coefficient as a
function of the Richardson number is shown in Fig. 2. As
expected, $D$ rapidly decreases with increasing Richardson number.\\

\begin{figure}
\plotone{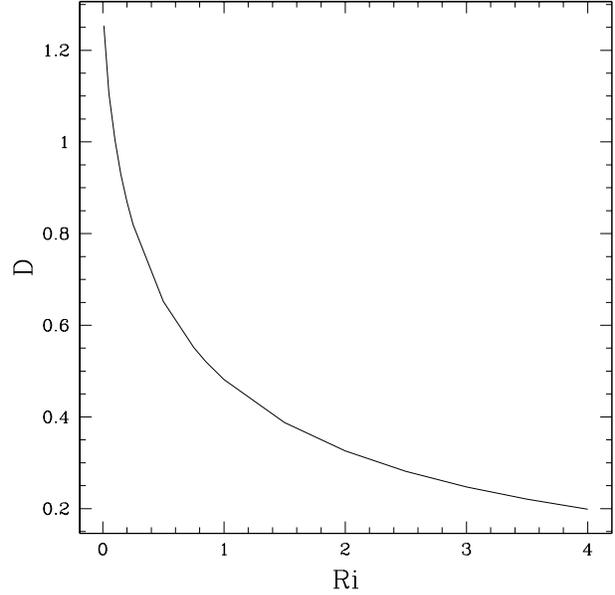}
\caption{Diffusion coefficient (in units of $U_0d$) as a function of the Richardson number.}
\end{figure}

In this section we presented an example of just one of the many
instability analyses found in the literature.  The example was chosen
for its simplicity. Of course, more realistic velocity and density
distributions have been studied since (see e.g. Chandrasekhar 1961,
Howard \& Maslowe 1973) but qualitatively they all yield the same
picture as shown in Fig. 1. Linear stability analyses, however, are
only valid for small perturbations.  They can be useful for studying
the onset of instability, but they are not very reliable in describing
the form or the efficiency of mixing itself. This is a problem that
requires numerical simulations and a set of such simulations will be
presented below.

In the next subsection we will quote a thermodynamic estimate of the
diffusion coefficient, before we move on to the numerical simulations
in Sec. 3.

\subsection{Maeder's prescription}

Maeder (1995, 1997) reexamined the Richardson criterion taking into
account radiative losses (also see Maeder \& Zahn 1998). These authors
derive a diffusion coefficient given by

\begin{equation}
D = K\frac{(dU/dz)^2}{N^2} ,
\end{equation}
where $K$ is the thermal diffusivity $K=4acT^3/3\kappa\rho^2C_P$, $a$
being the radiation density constant, $c$ the speed of light, $C_P$
the specific heat at constant pressure and $\kappa$ the opacity.
Maeder's prescription for mixing due to differential rotation has been
used, e.g., by Langer (1992) and Denissenkov (1994) for evolutionary
calculations for O and B stars in our Galaxy and the Large Magellanic
Cloud. More recently, Denissenkov \& Tout (2000) invoked a diffusion
constant of the form of Eq. (11) to explain extra mixing in globular
cluster red giants. 

\section{Numerical simulations}

Here we present the results of numerical simulations of the dynamical
shear instability in a stratified fluid.  The simulations were
obtained using the ZEUS-3D code which was developed especially for
problems in astrophysical hydrodynamics (Clarke \& Norman 1994).  The
code uses finite differencing on a Eulerian or pseudo-Lagrangian grid
and is fully explicit in time. It is based on an operator-split scheme
with piecewise linear functions for the fundamental variables. The
fluid is advected through a mesh using the upwind, monotonic
interpolation scheme of van Leer. For a detailed description of the
algorithms and their numerical implementation see Stone \& Norman
(1992a, b).\\

In our simulations we employed an ideal gas equation of state, we
ignored the effects of magnetic fields, rotation, nuclear reactions
and variations in radiative processes.  The simulations were computed
on a Cartesian grid and the computational domain was chosen to have
the dimension $2\cdot 10^8$ cm $\times$ $2\cdot 10^8$ cm $\times$
$10^8$ cm (in the $x$-, $y$- and $z$-direction, where gravity acts in
the $z$-direction).  It was covered by 100 $\times$ 100 $\times$ 50 grid
points.  The calculations were performed on a CRAY Jedi
parallel-processor and an IBM RS/6000 cluster.\\

\begin{figure}
\plotone{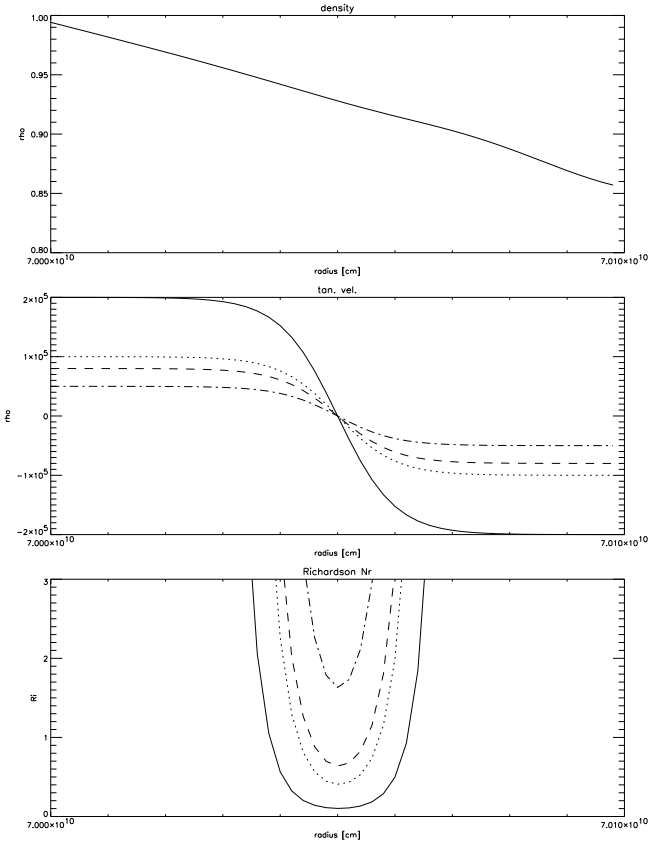}
\caption{Top: Initial density stratification; Middle: Shear velocities
for some selected simulations; Bottom: Corresponding Richardson numbers.}
\end{figure}

The gravitational acceleration was taken to be that of a point mass of
1 $M_{\odot}$ at a distance of 1 $R_{\odot}$ from the lower
boundary, where the gravitational acceleration acted purely in the
$z$-direction.  An analytic model for an isothermal density
distribution under constant gravitational acceleration was relaxed in
1D until hydrostatic equilibrium was attained. The residual velocities
after the relaxation were six orders of magnitude smaller than the
maximum shear velocities. The initial density distribution is shown in
Fig. 3.  Then a shear velocity profile was imposed on the fluid. It
was chosen to have the form of a hyperbolic tangent in order to
minimise the effect of the boundaries onto the shear layer, i.e.

\begin{equation}
U(z) = U_0\tanh [(z-z_0)/h] ,
\end{equation}
where $U_0$ is the amplitude of the shear velocity, $z$ the vertical
position of the shear layer, and $h$ its extent. In order to keep the
shear layer away from the boundaries, $h$ was taken to be smaller than
the vertical extent of the simulation region. Finally, $U_0$ was
chosen to yield a range of initial Richardson numbers of 0.05 - 3,
where the Richardson number is taken in its original simple
definition, i.e. Ri$=g\rho'/\rho {U'}^2$, and is measured at $z=z_0$.
Fig. 3 shows the initial conditions for Ri = 0.1, 0.4, 0.6 and
1.6. The boundary conditions were chosen to be periodic in the $x$ and
$y$ direction and reflecting in the $z$ direction.\\

\begin{figure}
\plotone{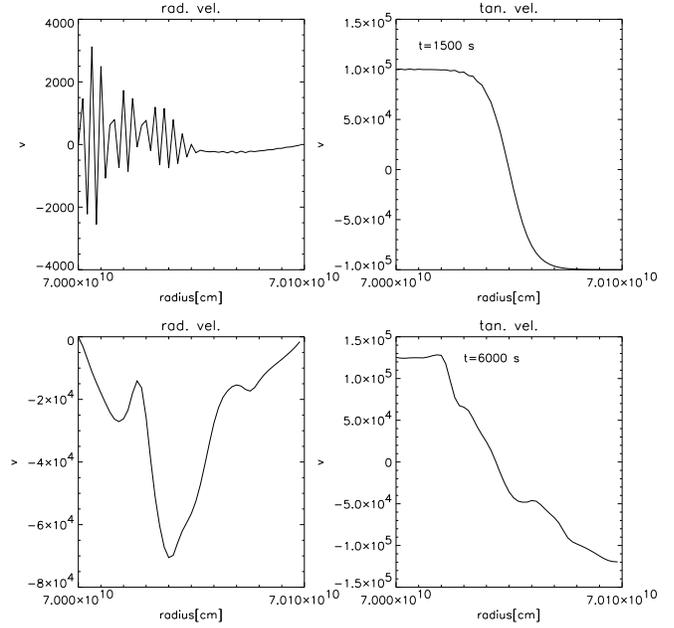}
\caption{Radial and shear velocities at various times for the example
with Ri=0.4}
\label{multi}
\end{figure}

\begin{figure}
\plotthree{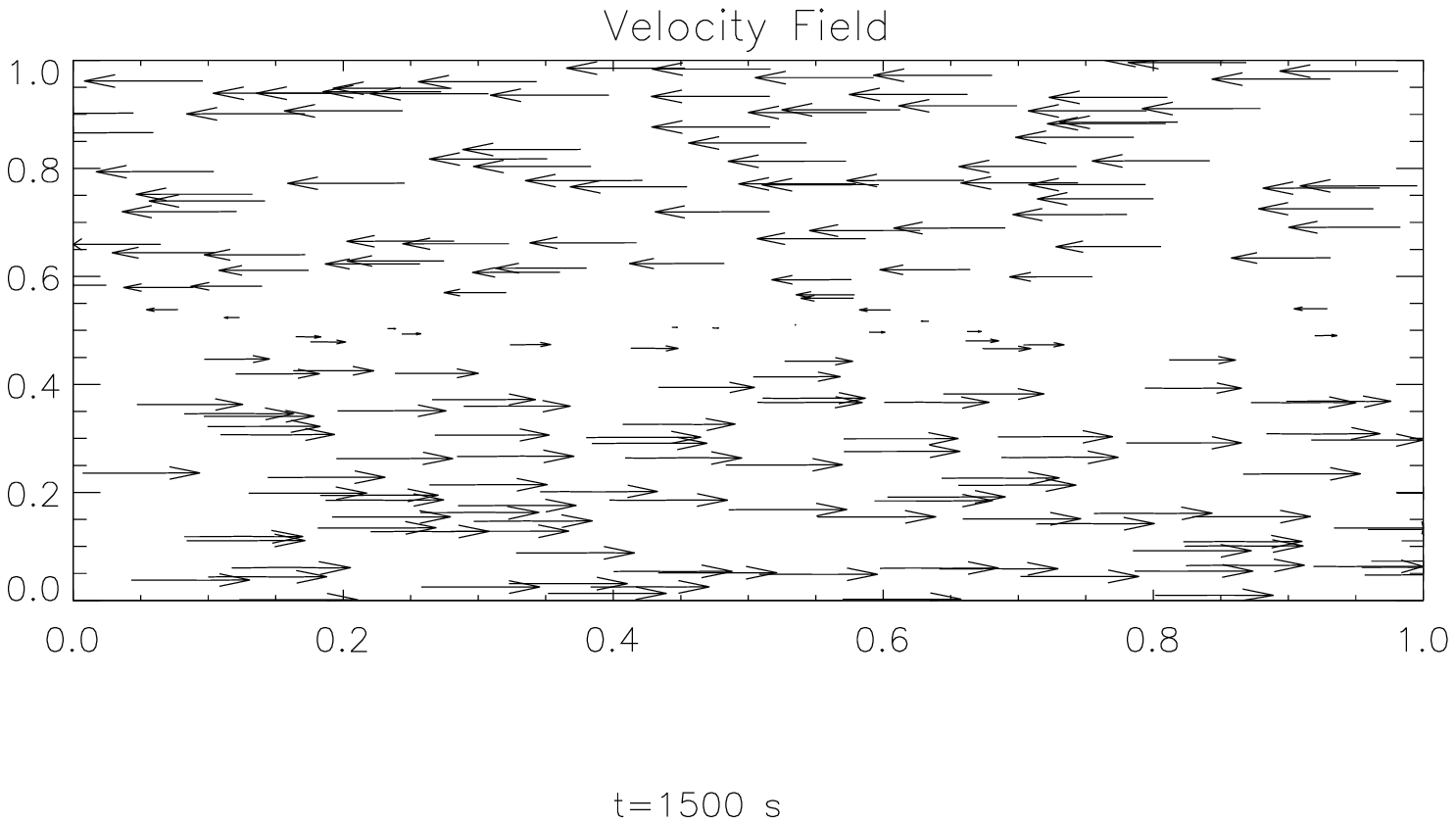}{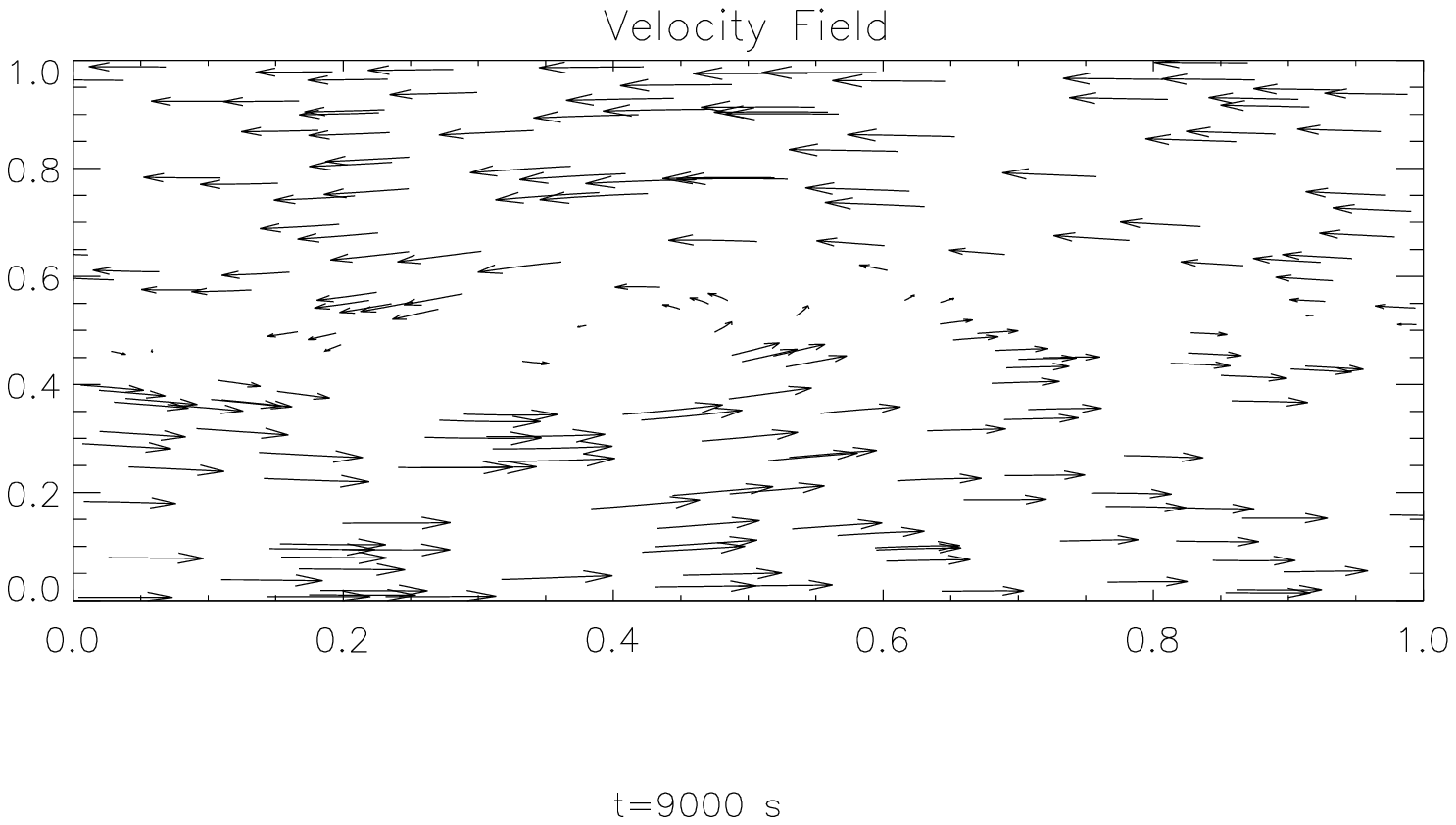}{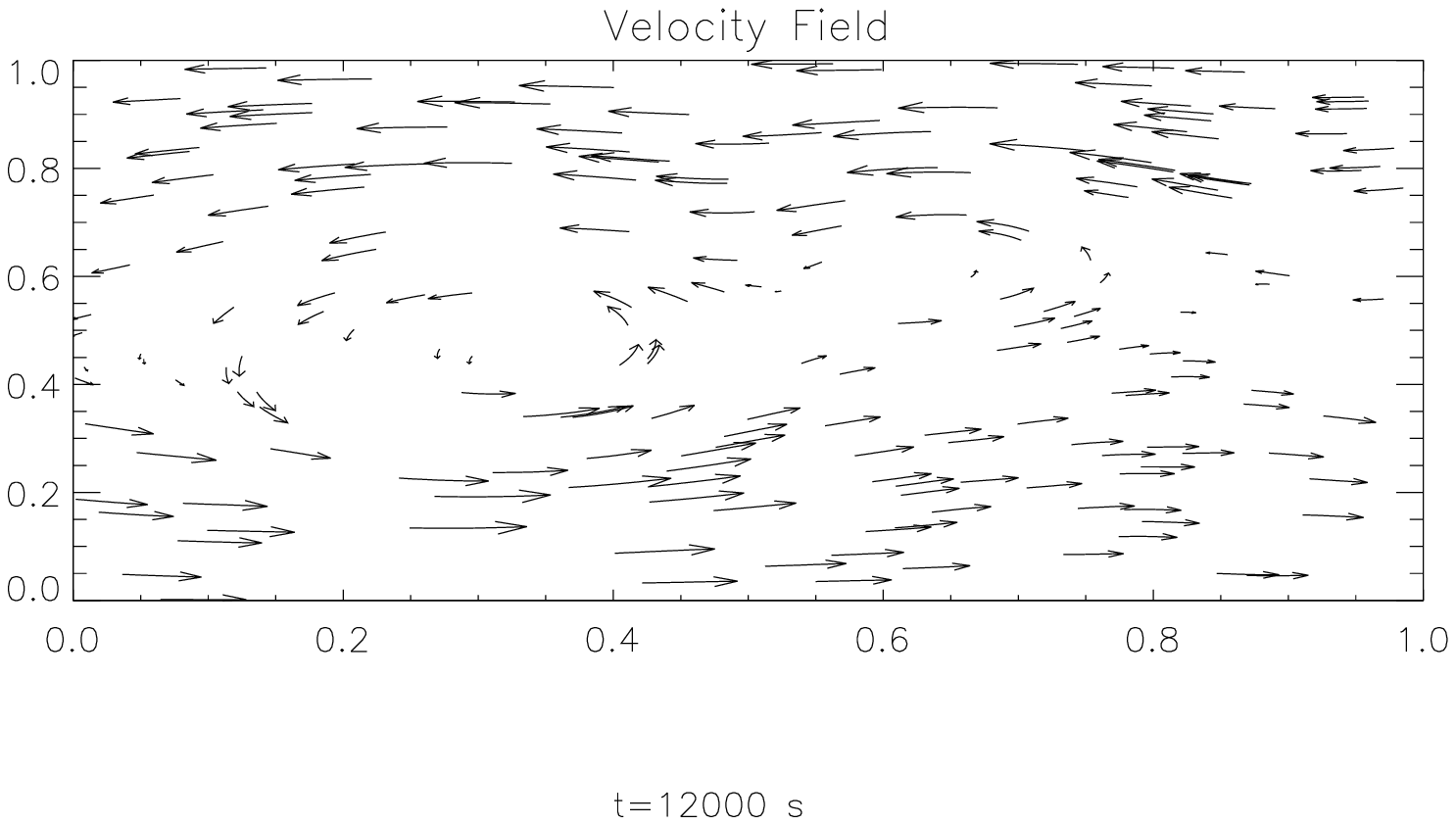}
\caption{Velocity field at $t=1500$ s, $t=9000$ s and $t=12000$ s for the same case as shown in Fig.\ref{multi}.}
\label{velo1}
\end{figure}

Figs. 4 and 5 show the $x$- and $z$-velocities in a vertical slice through
the computational domain at various times, for a flow with an initial
Richardson number of 0.4 in the shear layer.  The occurrence of
turbulence is clearly seen.  One can see that early on in the
simulation the flow undulates slightly before the instability becomes
fully nonlinear. Then one can observe that two vortices form which mix
material over a large section of the computational domain. The
vortices remain quite stable and increase in size. Their centres
coincide with the centre of the shear layer. One can also note that at
the centre of the vortex, the flow velocity is very small.  For
efficient transport the velocity fluctuations in the horizontal and
the vertical have to be correlated. This correlation is apparent in
the formation of vortices as seen in Fig. 5. More quantitatively it
can be expressed through the vorticity. Maps of the vorticity in the
direction both perpendicular to the gravitational acceleration and the
initial flow are shown in Fig. 6. (for Ri=0.4). The contours in
Fig. 6a. which show the vorticity after $t=1500$ s span a range of
$3\cdot 10^{-3}$ s$^{-1}$ compared to a range of $10^{-2}$ s$^{-1}$ in
Fig. 6b. after 9000 s. This example shows the production of vorticity
by the shear instability.\\

\begin{figure}
\plottwo{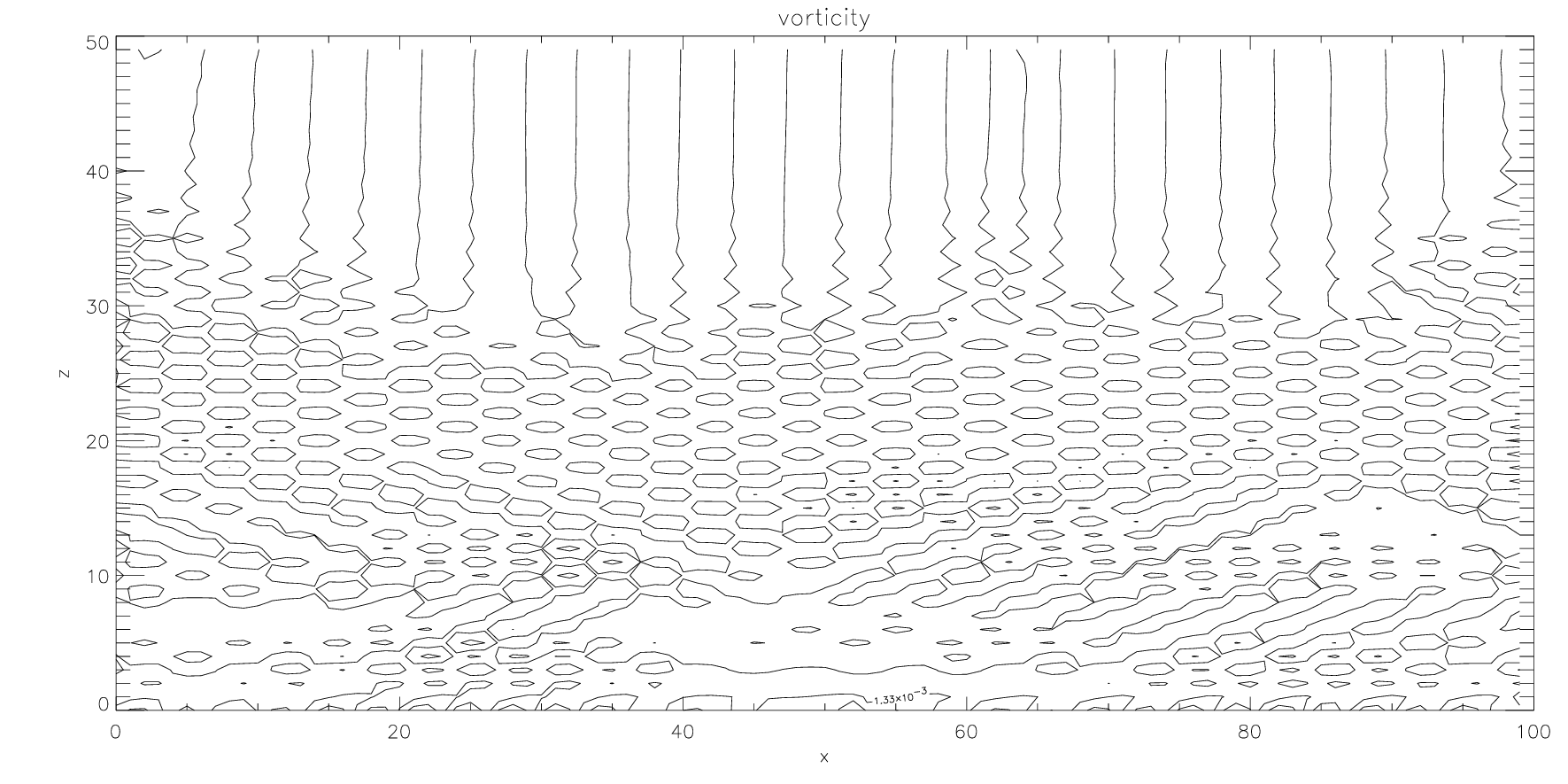}{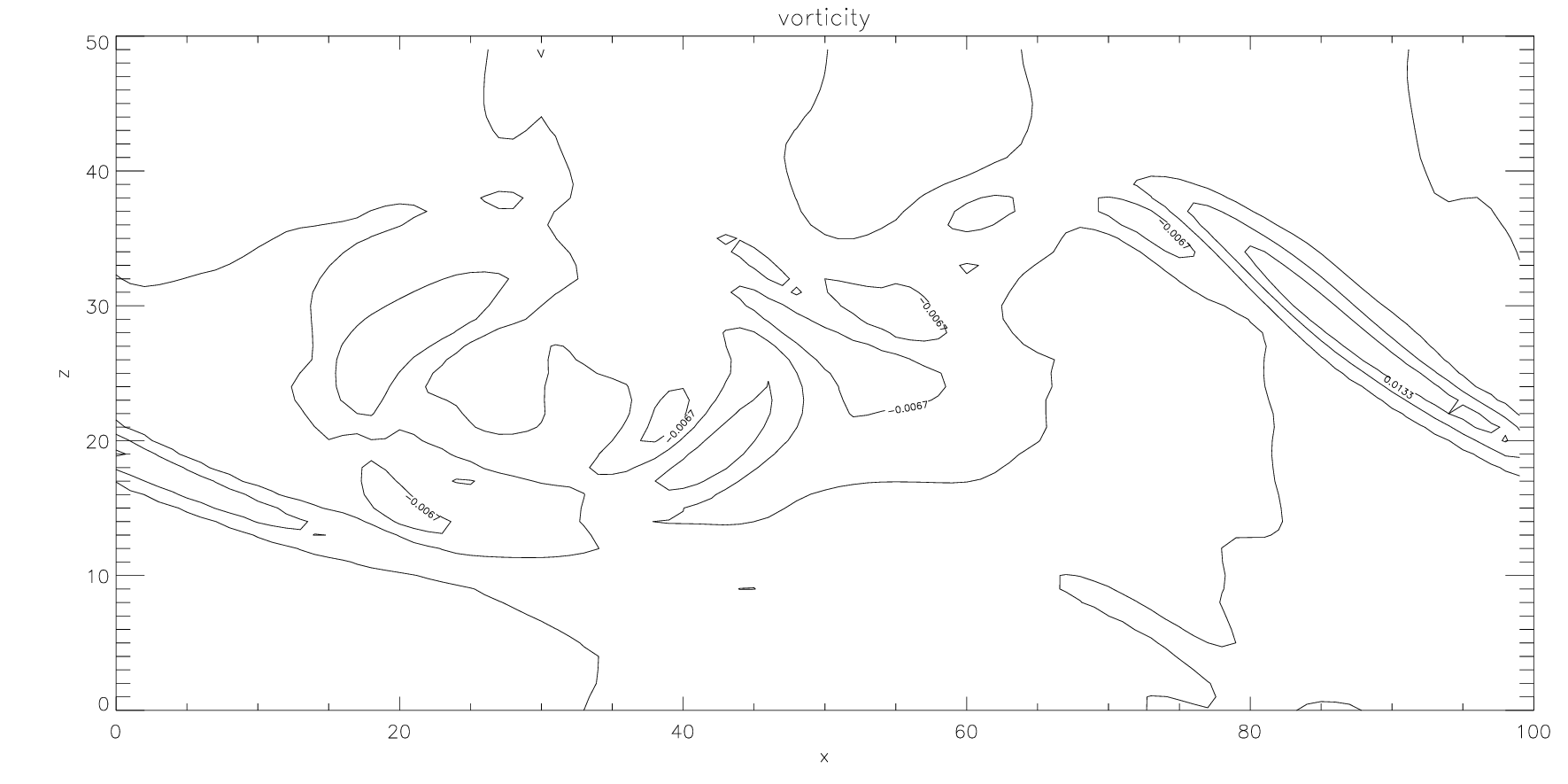}
\caption{Vorticity after $t=4500$ s and $t=12000$ s for the example
shown in Figs. \ref{multi}. In the top panel the contours span a range in vorticity of
$3\cdot 10^{-3}$ s$^{-1}$ and in the bottom panel a range of $10^{-2}$ s$^{-1}$.}
\end{figure}

\begin{figure}
\plotone{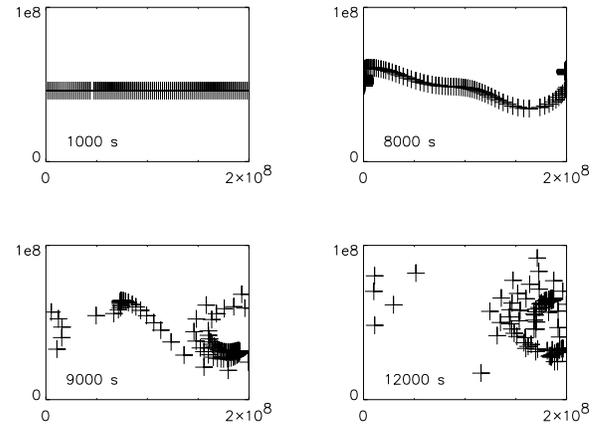}
\caption{Positions of the tracer particles for the example
shown in Figs. \ref{multi}.}
\end{figure}

The eddies have the form of long ``rolls'' which are aligned with the
$y$-axis, i.e. perpendicular to, both, the shear flow and the
gravitational acceleration.  They are found to stay fairly symmetric
in the $y$-direction through the entire depth of the computational
box. The structures on the big scales, as for example the ``rolls'' in
our simulations, `feel' the symmetry of the background model in the
$y$-direction and therefore, they retain this symmetry.  This has also
been observed in laboratory experiments with incompressible
fluids. Only on the small scales homogeneous, isotropic 3D turbulence
will develop but we do not resolve these scales in our simulations.
In the case of the shear instability, mixing is provided primarily by
motions on big scales and, consequently, we argue that for our purpose
we do not have to resolve these small scales. A further discussion of
this issue can be found in Sec. 3.2.

Furthermore, over longer distances the ``rolls'' will start to feel
the curvature of the star which is not accounted for in the slab
geometry of our simulations and they may have the effect of breaking
up the symmetry in the $y$-direction.  These issues remain
uncertainties in the results presented here and will have to be
examined in a later study.\\

\begin{figure}
\plotone{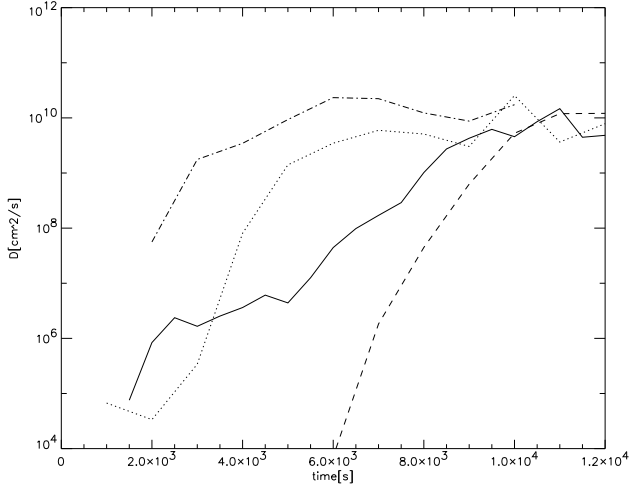}
\caption{Diffusion coefficient as a function of time for Ri=0.4. The solid line
shows the result of the 3D simulation and the dashed line the result
of the 2D simulation with the same resolution. The dotted line
represents the results from the computation on a 200 $\times$ 100 grid
and the dash-dotted line on a 400 $\times$ 200 grid.}
\end{figure}

In order to study mixing processes the ZEUS code was modified to
follow the motion of 1000 `tracer' particles which are advected with the
fluid. The positions of some of the tracer particles at various times (again
for the case with Ri = 0.4) are shown in Fig. 7.

The diffusion constant can then be defined as 

\begin{equation} 
D = \sigma^2/t ,
\end{equation}
where $\sigma^2=\frac{1}{N}\sum_N [z(N)-z_0]^2$, $z_0$ being the
original height of the $N$th tracer particle and $z(N)$ its height
after a time $t$.  As apparent from Fig. 7 the diffusion coefficient
varies with time.  The diffusion constant as a function of time
have been plotted in Fig. 8. In Fig. 8 the Richardson number is 0.4,
which is greater than 0.25 and, therefore, the flow should be stable
to the shear instability according to the simple Richardson
criterion. The simulations reveal that it is not, as has in fact been
shown in previous simulations.  One can observe that initially $D$
rises with time before it eventually approaches a value which remains
nearly constant for some time. The remaining scatter is mainly due to
the stochastic nature of the turbulent mixing. The transient phase
during which $D$ rises is longer, the greater the Richardson
number. Eventually, dissipation (mainly numerical) becomes noticeable
and $D$ starts to slowly decrease again.  The constant value of $D$ to
which the curves are converging, is the value that is of greatest
interest for the purpose of constructing stellar models. This value
for $D$ depends on the Richardson number. An understanding of the
quantitative dependence of $D$ on parameters such as the Richardson
number will be useful for constructing stellar models which treat
mixing through a diffusion equation.\\

We have plotted this `stationary' value of $D$ as a function of Ri for
all the simulations which we have performed (see Fig. 10). The
errorbars indicate the residual scatter observed in the
simulations.  Configurations with Ri $>$ 1.6 proved to be stable for
times up to 30,000 s.

\subsection{2D versus 3D simulations}

Two-dimensional simulations of convective instabilities are known to
introduce numerical artefacts and to yield unrealistic results. In
2D simulations of convection the energy cascades to
larger scales where viscous dissipation is less effective, whereas 3D
simulations reveal that the energy cascades to smaller scales where it
is absorbed. Moreover, 2D simulations often produce `rolls',
i.e. vortices in the direction perpendicular to the simulated
domain, which, for convection, tend to overestimate the mixing.\\

In order to investigate the importance of performing these simulations
in three dimensions, the runs described above were repeated in two
dimensions. The ZEUS code is designed to efficiently compute on 1- and
2-dimensional grids, so the runs could be readily repeated in 2D. The
same initial model and the same set of physical parameters as in the
3D simulations were used, and a grid of 100 $\times$ 50 points was
chosen.  Fig. 9 shows a sequence of snapshots of the velocity.\\

\begin{figure}
\plotthree{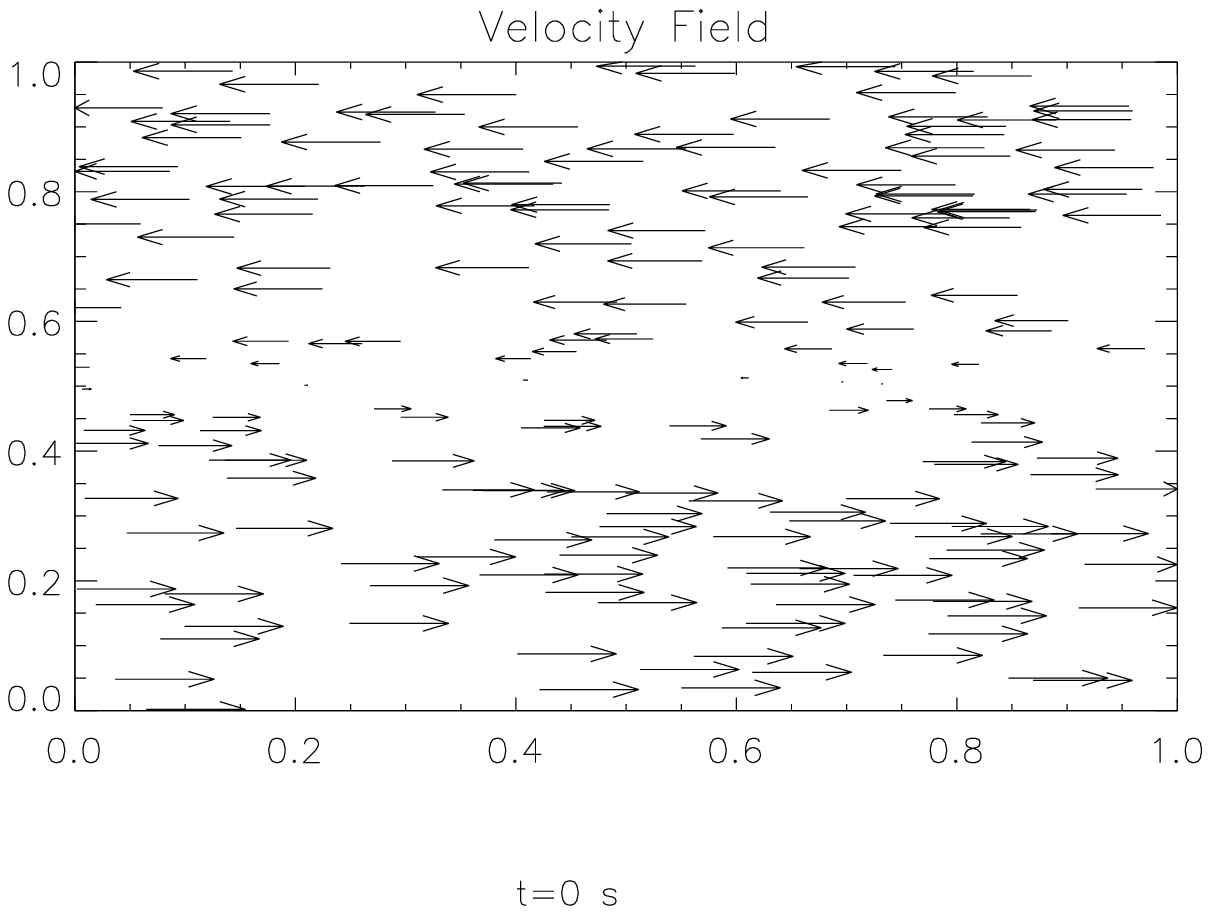}{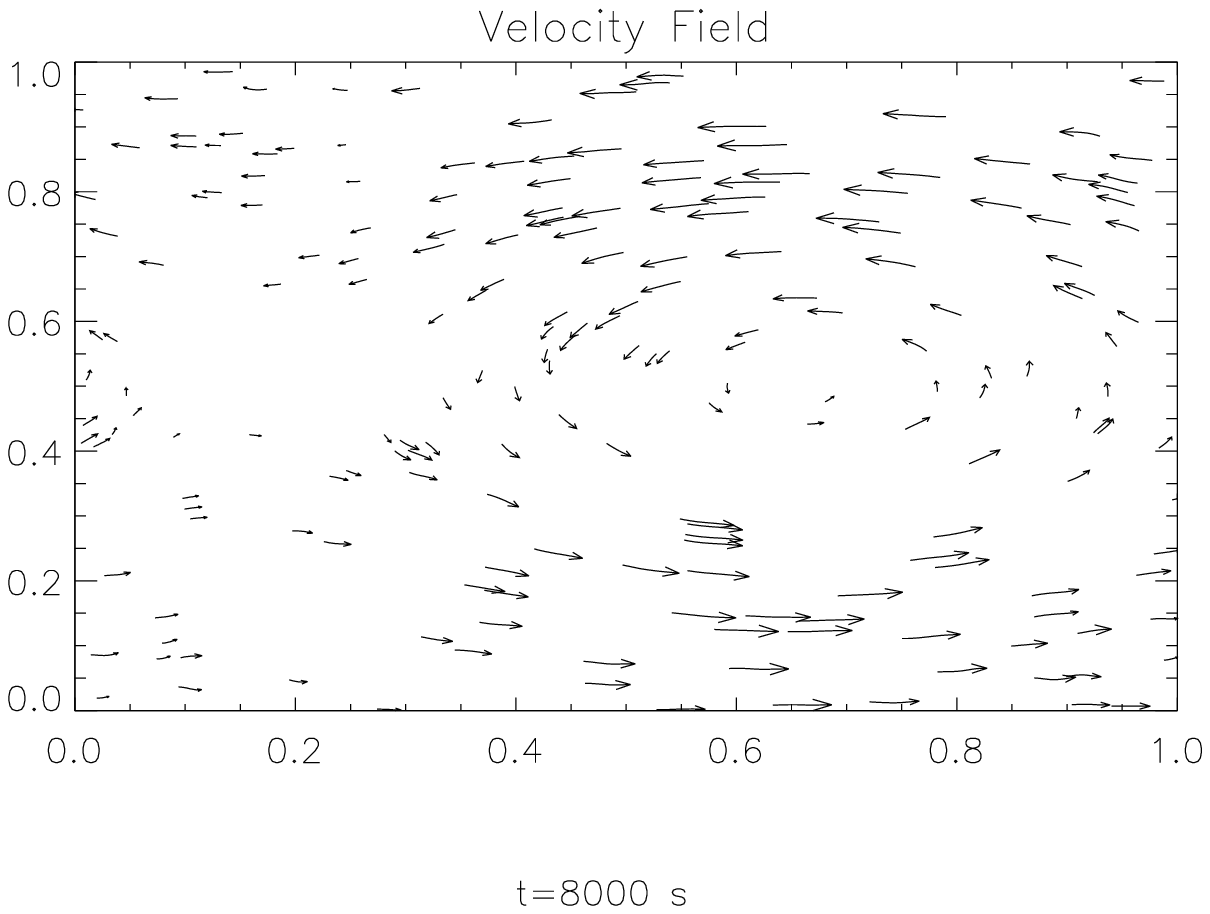}{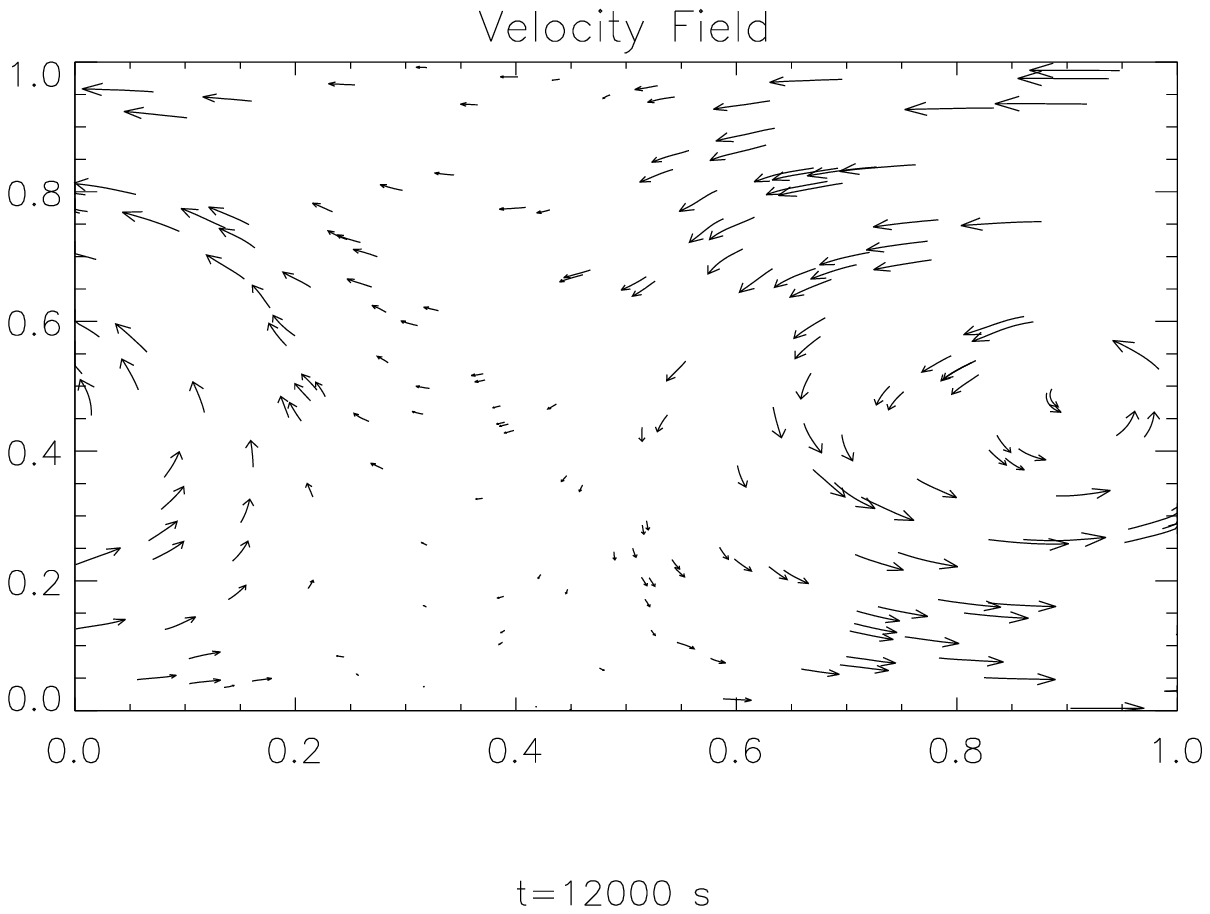}
\caption{Velocity fields in a 2D simulation after $t=0$ s, $t=8000$
s and $t=12000$ s.}
\end{figure}

\begin{figure}
\plotone{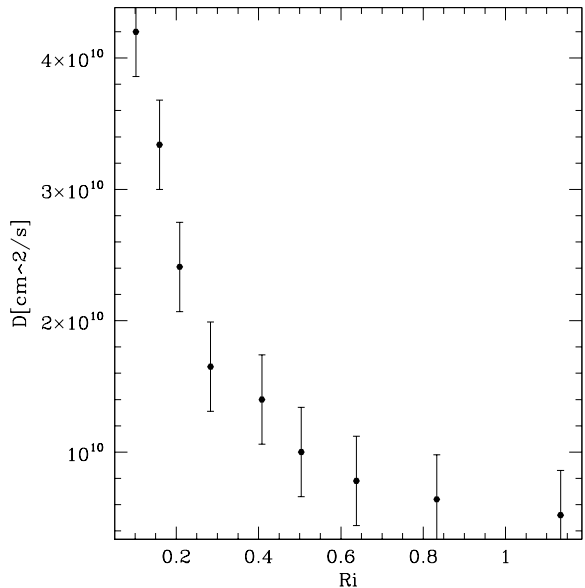}
\caption{Diffusion coefficients for different Richardson numbers.}
\end{figure}

The diffusion coefficient derived from the 2D simulations as a
function of time is displayed as a dashed line in Fig. 8.  The figures
show that in 2D simulations $D$ rises more slowly with time than in
three dimensions.  But eventually $D$ converges to about the same
value which was obtained in the 3D simulations. This is an encouraging
result. Not only does this raise the trust in the credibility and
self-consistency of these results, it also suggests that 2D
simulations may be sufficient to determine the efficiency of shear
mixing. This conclusion may not be too surprising given that the 3D
simulations revealed a flow structure consisting of long rolls that
are parallel to the $y$-axis, i.e. perpendicular to the shear flow.
On smaller scales, near the Kolmogoroff scale, the picture would look
very different because one would observe the onset of full isotropic
3D turbulence.  But here we are only simulating the Kelvin-Helmholtz
instability and not full 3D turbulence which is a much more difficult
problem and which requires a much finer (and yet unattained) numerical
resolution. This will be discussed further in the next section.

\subsection{Numerical viscosity}

One frequently voiced objection to these kinds of direct numerical
simulations is that Reynolds numbers as high as those encountered in
stars are unattainable on current computers, and that therefore the
results are unrealistic. But, as pointed out by Balbus, Hawley \&
Stone (1996), this criticism is unjustified for simulations of the
shear instability. They argue, that in order to simulate the onset of
instability in a laminar flow, it is merely necessary that the
`typical' wavelength of the instability is resolved by the numerical
scheme and that the numerical diffusion at this wavelength is less
than the growth rate. This makes the simulation of shear instabilities
an easier task than the simulation of viscous instabilities where in
theory one would have to resolve everything down to the viscous length
scale.\\

If the flow is unbounded, there is no viscous boundary layer which
might interfere with the results. The nonlinear instabilities are
fundamentally inviscid in character. Therefore, we only need a
resolution capable of resolving a range of wavelengths with numerical
diffusion errors less than the growth rates.  This view is also
supported by experimental observations which suggest that mixing
through shear instabilities is dominated by large scale motions.\\

In Fig. 12 we have plotted the kinetic energy in the vertical
($z$-direction) as a function of time. The vertical kinetic energy is
expressed in units of the total initial kinetic energy. In the
beginning the energy is very small until the instability sets in. Then
the kinetic energy jumps to about 10\% of its total initial value and
slowly decreases amidst big fluctuations. This decrease is mainly due
to numerical viscosity and shows yet again that we are not simulating
proper 3D turbulence which would show a saturation of the kinetic
energy.\\

Porter \& Woodward (1994) have estimated the Reynolds number of
hydrodynamical simulations based on a PPM (piecewise parabolic method)
code. The Reynolds number depends on the truncation error of the
finite-difference algorithm, the Courant number, the background
advection and the number of grid points. They found that the effective
Reynolds number is proportional to the third power of the number of
grid points, with the main dissipation occurring at short
wavelengths. Above this critical wavelength the diffusion was found to
be small. Since the ZEUS code uses piecewise linear functions instead
of piecewise parabolic functions, the truncation errors of ZEUS will
be larger than those of a PPM code. But even if they were only
proportional to the second power of the number of grid points in one
direction, we would still expect a numerical Reynolds number of around
$10^4$.

The effect of the numerical viscosity is to suppress fluctuations
beneath the `effective' viscous length scale. This will become
important for high enough Richardson numbers and all numerical
simulations will find stability above a certain Richardson number. In
our case the flow remained stable for Ri $>$1.6. Higher resolutions, and
therefore higher Reynolds numbers will push this limiting Richardson
number to slightly higher values.

\subsection{Dependence on grid size}

We have investigated the dependence of our results on the resolution
of the computational grid by repeating the 2D runs presented above
(100 $\times$ 50) on grids with (b) 200 $\times$ 100 and (c) 400
$\times$ 200 grid points. The variation of $D$ with time is shown in
Fig. 8. by the dotted (b) and dash-dotted (c) lines. It is immediately
evident that the behaviour of $D$ with time varies with the numerical
resolution. For more finely resolved grids, $D$ rises more
quickly. This was to be expected since higher resolutions produce
higher Reynolds numbers and therefore a less viscous flow. For higher
effective Reynolds numbers the turbulence grows more rapidly, and this
is observed here. But again it merits mention that for all simulations
$D$ eventually converges to rougly the same value for a given
Richardson number, as long as the Richardson number is not too
big. For large Richardson numbers, the numerical viscosity becomes
important, but here we will restrict the discussion to Richardson
numbers less than 2.\\

Finally, it can be seen in Fig. 5 that the eddies nearly span the entire
vertical extent of the simulation domain.  Hence, it is interesting
to verify whether the vertical size of the computational domain
affects the size of the eddies, and therewith the value of the
mixing coefficients.  We have repeated some 2D simulations in
computational domains which had a vertical extent of more than $50 h$
or ten times the vertical size of the previously shown simulations. We
did not find that the eddies became bigger when the size of the
computational box was increased.

\section{Discussion}

In this paper results of direct numerical simulations of shear
instabilities were presented.  The simulations were carried out using
the ZEUS hydro code. A chemically homogeneous stratified fluid in
hydrostatic equilibrium was set up, onto which a shear flow was
superimposed. The mixing was quantified by studying the dispersion of
`tracer' particles that are advected by the fluid.  Thus, the
variation of the diffusion constant with the Richardson number was
examined.\\

In accordance with previous simulations, it was found that efficient
mixing occurs for Richardson numbers substantially higher than 0.25,
contrary to the simple linear stability criterion.  Now the diffusion
coefficients that were derived from the motion of the tracer particles
can be compared to the analytical estimates discussed in Sec. 2.
Applying Maeder's formalism to our initial model, i.e. Eqn (11),
yields substantially lower values for $D$ than found in our
simulations.  For our initial models, Maeder's diffusion coefficients
in the centre of the shear layer lie between values of $10^6$ - $10^7$
cm$^2/$s.  This is more than 3 orders of magnitude smaller than the
diffusion coefficients found here.  Finally, one might want to compare
our results with Fig. 2, even though this figure assumed a different
shear velocity distribution than the hyperbolic tangens used in the
simulations.  In Fig. 11 where the dimensionless diffusion coefficient
(in units of $U_0 h$) is plotted against the Richardson number. This
dimensionless presentation is also useful for a wider application of
our results.  It is interesting to note that now our values for $D$
are smaller by almost two orders of magnitude than those predicted by
the linear analysis of our simple model of Sec. 2.\\

\begin{figure}
\plotone{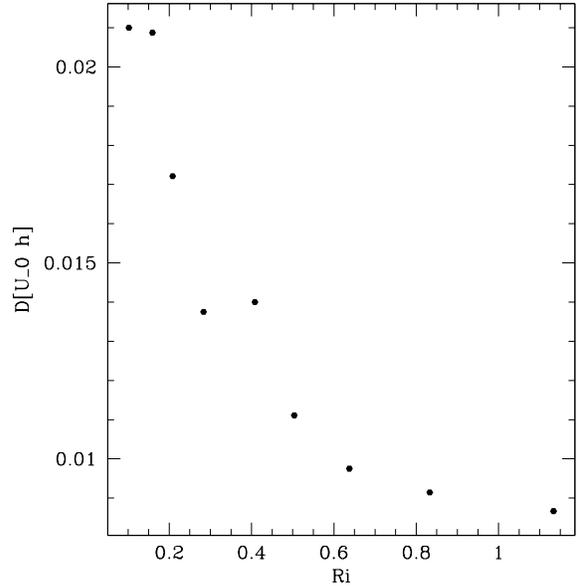}
\caption{Diffusion coefficients for different Richardson numbers in
units of $U_0 h$ (errorbars have been omitted).}
\end{figure}

It was argued that numerical viscosity has no great effect on our
findings as the effective Reynolds number is small on the scales on
which the Kelvin-Helmholtz instability operates. This is true as long
as the Richardson number is not too big.  Furthermore, the differences
between 2D and 3D simulations of shear instabilities were studied. We
found that the dimensionality of the simulation hardly affects the
final `steady state' diffusion coefficient. It only affects the
dynamics of the onset of the instability until a quasi-stationary
state has been reached. The same qualitative observation was made when
the resolution of the numerical grid was varied. It was found that the
resolution affects the initial temporal variation of the diffusion
coefficient, but hardly affects the final value of $D$.  However, the
similarity between the 2D and 3D simulations is only valid on the big
scales which we are resolving. On smaller scales, near the Kolmogoroff
scale, one would observe full isotropic 3D turbulence which looks very
different in two dimensions.\\

To make the Richardson number the only controlled parameter in our
study, is clearly a simplification of the factors that
determine the efficiency of mixing. In reality, the efficiency of
mixing will depend on the density stratification and the velocity
gradient separately, and not solely on the Richardson
number. Moreover, the mixing will depend on the exact shape of the
velocity profile and, only to a first approximation, on its first
derivative. Nevertheless, if one is interested in the efficiency of
mixing as a function of a single parameter, e.g. for stellar evolution
studies, the Richardson number is still the most suitable parameter,
since in the limit of an inviscid and incompressible shear flow, the
stability is governed by the Richardson number.  \\

In this work the effects of the dynamical shear instability in a plane
shear layer were investigated. Clearly, this is not the only mechanism
responsible for mixing in stars. In rotating fluid bodies other
instabilities may occur, such as, e.g., the Solberg-Holand
instability, the Eddington-Sweet instability and the
Goldreich-Schubert-Fricke instability. However, the shear instability
operates on the dynamical timescale whereas the instabilities tied to
the rotation occur on the longer Eddington-Vogt timescale. Therefore,
these instabilities operate on a much longer timescale and therefore
play a lesser role in diffusion processes, even though they can affect
the long-term evolution of the star.\\

\begin{figure}
\plotone{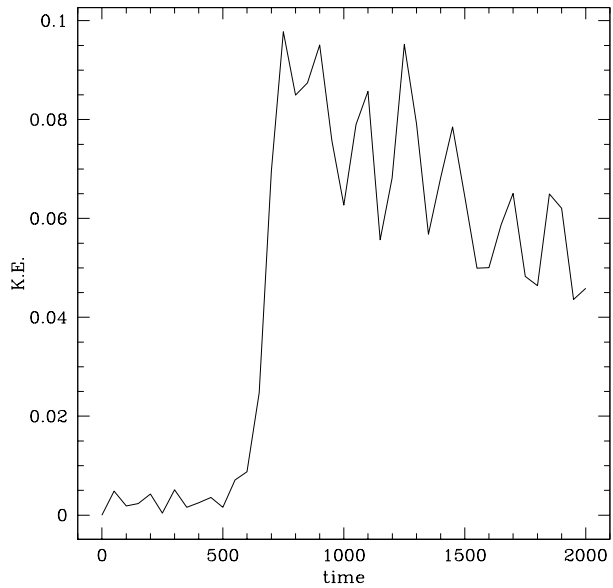}
\caption{Kinetic energy in the vertical $z$-direction as a function of
time. The kinetic energy is expressed in units of the total initial
kinetic energy and the units of time are arbitrary.}
\end{figure}

The consequences of the newly found diffusion coefficients onto
stellar evolution and surface abundances are difficult to foresee. In
stars the speed as well as the depth of mixing determine the balance
between mixing and nuclear burning. Therefore, the effect of mixing
depends sensitively on the conditions prevailing in the star, as for
example the positions and extents of its convection and burning
shells.\\

Finally we should mention that a number of factors can inhibit or
facilitate mixing such as gradients in the chemical potential of the
fluid, diffusion of radiation, magnetic fields and effects pertaining
to the spherical geometry.  In late-type stars strong chemical
composition gradients exist, which will have a stabilising effect on
the stratification. Therefore, especially for stellar evolution
studies, the chemical composition gradient is an important parameter,
which will have to be included in future work.  Similarly, poloidal
magnetic fields, as they are believed to exist in the Sun, will also
suppress mixing processes. These effects are potentially of great
importance in stars and will be studied in a forthcoming paper.

\section*{Acknowledgements}

A part of the simulations were performed on computers of the
Rechenzentrum Garching. We thank Phil Armitage for helpful
discussions and the referee for useful comments.


\begin{thebibliography}{99}

\bibitem{} Balbus, S.A., Hawley, J.F., \& Stone, J.M. 1996, \apj, 467,
76
\bibitem{} Balmforth, N.J., Morrison, P.J., 1998, Studies in Applied
Mathematics, preprint no. physics/9809024
\bibitem{Chand} Chandrasekhar, S., 1961, {\it Hydrodynamic and
Hydromagnetic Stability}, Clarendon Press, Oxford 
\bibitem{} Canuto, V.M., 1998, \apj, 508, 767
\bibitem{} Charbonnel, C. 1995, \apj, 453, L41
\bibitem{} Denissenkov, P.A. 1994, A\&A, 287, 113
\bibitem{} Denissenkov, P.A., Tout, C.A., 2000, MNRAS in press
\bibitem{} Endal, A.S., Sofia, S. 1978, \apj, 220, 279
\bibitem{} Fransson, C. et al. 1989, \apj, 336, 429
\bibitem{} Freytag, B., Ludwig, H.-G., Steffen, M. 1996, A\&A, 313, 497
\bibitem{Ha} Hazel, P. 1972, J. Fluid Mech., 51, 39
\bibitem{} Heger, A. 1996, {\it The Presupernova Evolution of Rotating
Massive Stars}, PhD thesis, Max-Planck-Institut f\"ur Astrophysik,
Garching
\bibitem{} Herrero, A. et al., 1992, A\&A, 261, 209
\bibitem{HoMa} Howard, L.N., Maslowe, S.A. 1973, Boundary-Layer Meteorology, 4, 511
\bibitem{KipTho} Kippenhahn, R., Thomas, H.-C., 1978, A\&A, 63, 265
\bibitem{} Kraft, R.P., 1994, PASP 106, 553
\bibitem{Kr} Kraft, R.P., Sneden, C., Smith, G.H., Shetrone, M.D.,
Langer, G.E., Pilachowski, C.A., 1997, AJ 113, 279
\bibitem{} Langer, N., 1992, A\&A, 265, L17
\bibitem{} Langer, N., Maeder, A., 1995, A\&A, 295, 685. 
\bibitem{} MacDonald, J., 1983, \apj  273, 289
\bibitem{Mae1} Maeder, A., 1997, A\&A, 321, 134
\bibitem{Mae2} Maeder, A., 1995, A\&A, 299, 84
\bibitem{Mae4} Maeder, A., Zahn, J.-P., 1998, A\&A, 334, 1000
\bibitem{} Maslowe, S.A., Thompson, J.M., 1971, Phys. Fluids 14, 453
\bibitem{Mae3} Meynet, G., Maeder, A., 1997, A\&A, 321, 465
\bibitem{MoGi} Morris, P.J., Giridharan, M.G., Lilley, G.M., 1990,
Proc. R. Soc. Lond. A, 431, 219
\bibitem{} Nordlund, A., Stein, R.F. 1996
\bibitem{} Pinsonneault, M.H., et al., 1991, \apj, 367, 239
\bibitem{HoMa} Porter, D.H., Woodward, P.R., 1994, ApJS, 93, 309
\bibitem{} Shu, F.H,  {\it The Physics of Astrophysics Vol. II}, 1992,
Univ Science, Mill Valley
\bibitem{} Singh, H. P., Roxburgh, I. W., Chan, K. L. 1998, A\&A, 340,
178
\bibitem{} Staritsin, E.I., 1999, Astron. Rep., 43, 592
\bibitem{NoSt1} Stone, J.M., Norman, M.L., 1992a, ApJS, 80, 753
\bibitem{NoSt2} Stone, J.M., Norman, M.L., 1992b, ApJS, 80, 791
\bibitem{} Venn, K.A., Lambert, D.L. \& Lemke, M. 1996, A\&A, 307, 894
\bibitem{} Zahn, J.-P. 1991, A\&A, 252, 179
\end{thebibliography}
\end{document}